\documentclass[12pt]{article}%
\usepackage{amsmath}
\usepackage{amsfonts}
\usepackage{amssymb}
\usepackage{graphicx}
\usepackage{epsf,color,colordvi,pifont}
\usepackage{amsfonts}%
\providecommand{\U}[1]{\protect\rule{.1in}{.1in}}
\begin{document}

\title{Comment on Cherenkov radiation in magnetized vacuum}
\author{{\large A.~E.~Shabad}\thanks{shabad@lpi.ru}\\\textit{P.~N.~Lebedev Physics Institute, Moscow, Russia} }
\maketitle

\begin{abstract}
The Cherenkov radiation in the vacuum with constant magnetic background is
denied within quantum kinematics in spite of the refractive index of that
equivalent medium being greater than unity.

\end{abstract}

\newpage

\section{Introduction}

It was noted rather long ago \cite{Shabad2004} that there is no Cherenkov
effect in the vacuum filled with a strong magnetic field, in spite of the fact
that the latter supplies it with nontrivial dielectric permeability
\cite{Erber} and hence with the possibility that a charged particle may
propagate with the speed exceeding that of light in this equivalent optically
dense medium. On the other hand, recently \cite{1}, the opposite was claimed,
and Cherenkov radiation (ChR) due to classical motion of a charge across the
magnetic field was estimated, the circular orbit being approximated by a
straight line for this purpose. The effect was also recognized in a number of
subsequent publications, e.g. \cite{Bulanov}, \cite{Lee}.

In the present Letter I reinforce the denial of ChR in the magnetized
vacuum\footnote{This objection does not relate to ChR in a plane-wave field,
which is another subject in \cite{1}, as well as in \cite{Dremin},
\cite{Bulanov}.}. I state that the full quantum treatment of the charge motion
provides a kinematic ban against ChR. Note that in the standard Cherenkov
effect the quantum treatment of its kinematics was proposed by V.L. Ginzburg
in 1940 \cite{Ginsburg}. The corresponding formulas are equivalent to the
classical ones derived in the pioneering work \cite{Tamm} on ChR only when the
energy of the irradiated photon and, correspondingly, the energy of the
irradiating charge are much smaller than its mass of rest \cite{Ginsburg1996}.
Evidently, interests of nowadays extend far beyond this restriction.

I shall refer to the quantum relativistic electron-positron gas in the
magnetic field that exhibits both the synchrotron and Cherenkov radiation
united into their common Compton scattering matrix element $e\rightarrow
e\gamma$. To separate these different types of radiation I define ChR as the
one that has its kinematic border at a certain speed of the charged particle.
I demonstrate that in the limit where the plasma is absent, ChR inherent to it
disappears, while the synchrotron radiation remains.


\section{Quantum kinematics of the process $e\rightarrow e\gamma$ and bans on
the Cherenkov radiation in vacuum with magnetic field}

In an external magnetic field the process of a photon emission by an electron
$e\rightarrow e\gamma$ exists on the mass shell of the involved particles
already in the vacuum.

Let, in any reference frame, where only the magnetic part $\mathbf{B}$ \ of a
constant external field exists, the electron be characterized by its momentum
component $p_{\parallel}$ along $\mathbf{B}$ and by the Landau quantum number
$n$ presenting partially the transverse degree of freedom of the electron. The
energy of the electron is $\varepsilon(p_{\parallel},n)=\sqrt{m^{2}%
+p_{\parallel}^{2}+2enB,}$ where $m$ is the electron mass and $e$ its charge
($c=\hbar=1)$.

There is another continuous quantum number responsible for the center-of-orbit
coordinate in the plane ($X,Y$) orthogonal to $\mathbf{B}$ (the axis directed
along it is understood as $Z)$. Choosing the vector-potential of the external
field to be in the linear gauge $A_{x}=-BY,$ $A_{y}=A_{z}=0$ and directing the
$y$-axis so that it pass through the center of orbit, $X=0,$ $Y=Y_{0},$ we may
write the corresponding "momentum" component in the plane orthogonal to the
magnetic field as $p_{x}=Y_{0}eB$. The energy does not depend on this quantum
number, however.

Let the electron produce a photon with the energy $k_{0}$ and its momentum
components along and across the field $\mathbf{B}$\ being $k_{\parallel}$ and
$\mathbf{k}_{\perp}$, resp. Once the vector-potential is independent of time
and of the coordinate $X$ the conserving components of the energy-momentum are
the energy and the momenta along the magnetic field and $p_{x}$
\begin{equation}
p_{\parallel}=k_{\parallel}+p_{\parallel}^{\prime},\text{ \ \ \ \ }%
\varepsilon(p_{\parallel},n)=k_{0}+\varepsilon(p_{\parallel}^{\prime
},n^{\prime}),\text{ \ \ \ \ }p_{x}=\left(  \mathbf{k}_{\perp}\right)
_{x}+p_{x}^{\prime} \label{1}%
\end{equation}
Primes here mark the final state. The third conservation law here will not
play a role within the present consideration.

Calculation of matrix elements in \cite{2} shows a profound difference between
transitions with $n\neq n^{\prime}$ and $n=n^{\prime}$ in what concerns the
threshold behavior of the corresponding matrix elements: it is infinite in the
first case and finite in the second. It is natural to refer to processes with
$n\neq n^{\prime}$ as the synchrotron radiation and attribute the term "ChR"
to the process where electron emits a photon without changing its Landau
quantum number. The physical ground for this classification will become clear
later below.

\subsection{Radiation of a photon by electron without changing its Landau
quantum number, $n=n^{\prime}$}

Now I set $n=n^{\prime}$ and denote $m_{n}^{2}=m^{2}+2eB.$ By squaring
equations (\ref{1}) and solving them with respect to $p_{\parallel}$ we
obtain
\begin{equation}
p_{\parallel}^{(1,2)}=\frac{k_{\parallel}}{2}\pm\frac{k_{0}}{2}\sqrt
{1+\frac{4m_{n}^{2}}{k_{\parallel}^{2}-k_{0}^{2}}}. \label{p3(12)}%
\end{equation}
This is real in two regions. One is $k_{0}^{2}-k_{\parallel}^{2}%
\geqslant4m_{n}^{2}$. In this domain Eqs. (\ref{p3(12)}) are redundant
solutions acquired by the squaring. They correspond to the minus sign in front
of $\varepsilon(p_{\parallel}^{\prime},n^{\prime})$ in (\ref{1})\ and serve
the production by a photon of an electron-positron pair, which is not the
process under consideration. Therefore, we are left with the other possible
region
\begin{equation}
z\doteq k_{\parallel}^{2}-k_{0}^{2}\geqslant0,\text{ }z<k_{\parallel}^{2}.
\label{Domain}%
\end{equation}
Only this domain is kinematically admitted for the reaction considered with
$n=n^{\prime}$. However, even within this domain not the both values
(\ref{p3(12)}) are solutions to the set (\ref{1}) with $n=n^{\prime}$. To
select the genuine solutions, let us substitute (\ref{p3(12)}) back into
(\ref{1}). We obtain for the energies of the initial and final electron%
\begin{align*}
\varepsilon^{(1,2)}  &  =\sqrt{m_{n}^{2}+(p^{(1,2)})^{2}}=\left\vert
\frac{k_{0}}{2}\pm\frac{k_{\parallel}}{2}\sqrt{1+\frac{4m_{n}^{2}}{z}%
}\right\vert ,\\
\varepsilon^{\prime(1,2)}  &  =\sqrt{m_{n}^{2}+\left(  p_{\parallel}%
^{(1,2)}-k_{\parallel}\right)  ^{2}}=\left\vert -\frac{k_{0}}{2}\pm
\frac{k_{\parallel}}{2}\sqrt{1+\frac{4m_{n}^{2}}{z}}\right\vert .
\end{align*}
Since $k_{0}>0,$ we have to choose the upper sign if $k_{\parallel}>0$ and
lower sign if $k_{\parallel}<0$, and then the second equation in (\ref{1})
$\varepsilon(p_{\parallel},n)=k_{0}+\varepsilon(p_{\parallel}^{\prime},n)$ is
fulfilled for each of these two choices. Finally, the genuine solution to the
set (\ref{1}) in the domain (\ref{Domain}) is%
\[
p_{\parallel}=\frac{k_{\parallel}}{2}+\frac{k_{0}}{2}\sqrt{1+\frac{4m_{n}^{2}%
}{z}}\text{sgn}k_{\parallel},\ \ \ \varepsilon=\frac{k_{0}}{2}+\frac
{|k_{\parallel}|}{2}\sqrt{1+\frac{4m_{n}^{2}}{z}}%
\]%
\begin{equation}
p_{\parallel}^{\prime}=-\frac{k_{\parallel}}{2}+\frac{k_{0}}{2}\sqrt
{1+\frac{4m_{n}^{2}}{z}}\text{sgn}k_{\parallel},\text{ \ \ \ \ }%
\varepsilon^{\prime}=\frac{k_{0}}{2}+\frac{|k_{\parallel}|}{2}\sqrt
{1+\frac{4m_{n}^{2}}{z}} \label{finChargeMom}%
\end{equation}
It is seen that due to the inequality (\ref{Domain}), the emitted photon
momentum component $k_{\parallel}$ along the magnetic field is directed in the
same way as that of the emitting electron. \qquad

Introducing the photon emission angle $\theta$ respective to the magnetic
field via the relation $\cos\theta=\frac{k_{\parallel}}{\mid\mathbf{k}\mid},$
and the refraction index as $N(k_{0},\mathbf{k})=\frac{\mid\mathbf{k}\mid
}{k_{0}}$ (the phase velocity of the emitted photon is the inverse of it,
$v_{\text{ph}}=\frac{1}{N}$), let us consider the case $k_{\parallel}>0,$ i.e.
the photon emitted ahead of the electron under the angle $\pi/2>\theta
>-\pi/2.$ In this case, in the domain (\ref{Domain}) it holds that
$\infty>\frac{k_{\parallel}}{k_{0}}=N\cos\theta\geqslant1.$ For the speed of
the irradiating electron we have the following expression%
\begin{equation}
v_{\text{ch}}=\frac{p_{\parallel}}{\varepsilon}=\frac{k_{\parallel}+k_{0}%
\sqrt{1+\frac{4m_{n}^{2}}{z}}}{k_{0}+k_{\parallel}\sqrt{1+\frac{4m_{n}^{2}}%
{z}}}=\frac{\lambda N\cos\theta+1}{\lambda+N\cos\theta}, \label{Chspeed}%
\end{equation}
where $\lambda=\frac{\sqrt{z}}{\sqrt{z+4m_{n}^{2}}}<1.$\ It is readily seen
that this speed is certainly not larger than unity (the speed of light in the
vacuum), but not smaller than the phase speed of light $v_{\text{ph}}=\frac
{1}{N}$. The first claimed inequality $1\geqslant v_{\text{ch}}$ reads
$\lambda+N\cos\theta\geqslant\left(  \lambda N\cos\theta+1\right)  $, or
$\left(  1-\lambda\right)  N\cos\theta\geqslant\left(  1-\lambda\right)  $.
This relation is evident since $N\cos\theta\geqslant1,$ while $1-\lambda$ is
positive and can therefore be cancelled from it$.$ The second claimed
inequality $v_{\text{ch}}\geqslant v_{\text{ph}}$ reads $\left(  \lambda
N\cos\theta+1\right)  N\geqslant\lambda+N\cos\theta.$ This is definitely
fulfilled, too, since $N^{2}\cos\theta\geqslant1$ and $N\geqslant N\cos
\theta.$

The same conclusions are achieved for the backward radiation.

The above analysis reveals every feature of the Cherenkov effect, but it
realizes only if the reaction takes place in a material medium, but not in a
magnetized vacuum, though $N$ may be greater than unity there. The point is
that the produced Cherenkov photon should fit a dispersion law, but no
dispersion curve is admitted in domain (\ref{Domain}) by the relativistic
covariance and the causality principle as far as the vacuum with a constant
magnetic field is concerned.$\qquad$

\subsection{Incompatibility with Lorentz covariance and causality}

In the vacuum filled with a constant external field there exist only two
independent Lorenz invariants that carry dependence on the photon momentum.
These are $k^{\mu}F_{\mu\nu}^{2}k^{\nu}$ and $k^{\mu}\widetilde{F}_{\mu\nu
}^{2}k^{\nu}$\ (The other invariant $k^{2}=$ $k_{0}^{2}-\mathbf{k}^{2}$ is
related to the previous two as $k^{\mu}\widetilde{F}_{\mu\nu}^{2}k^{\nu
}+k^{\mu}F_{\mu\nu}^{2}k^{\nu}=2\mathcal{F}k^{2}$). Here $F_{\mu\nu}^{2}$ and
$\widetilde{F}_{\mu\nu}^{2}$ are matrix squares of the external field tensor
and of its dual, and the only nonvanishing invariant of the external field is
denoted as $\mathcal{F=}\left(  F^{2}\right)  _{\mu}^{\mu}$. In any of the
special Lorentz frames where the external field does not contain an electric
part, the invariants become $\mathcal{F=}\frac{B^{2}}{2},$ $\frac{k^{\mu
}F_{\mu\nu}^{2}k^{\nu}}{2\mathcal{F}}=-k_{\perp}^{2},$ $\frac{k^{\mu
}\widetilde{F}_{\mu\nu}^{2}k^{\nu}}{2\mathcal{F}}=k_{0}^{2}-k_{\parallel}%
^{2}.$ The dispersion laws for each of the three photon polarization modes,
numbered by $i$, are to be found as solutions to the equations $k^{2}%
=\kappa_{i},$ where the Lorentz scalars $\kappa_{i}$ are eigenvalues of the
polarization $4\times4$ tensor \cite{Batalin}, \cite{Baier}. They depend on
the above invariants. Hence in the special frames the photon dispersion laws
may be written in the form
\begin{equation}
k_{0}^{2}-k_{\parallel}^{2}=f_{i}(k_{\perp}^{2},B). \label{dspr}%
\end{equation}
We shall omit an explicit indication of the argument $B$ in what follows. It
follows from (\ref{dspr}) that the refraction index $N=\frac{|\mathbf{k}%
|}{k_{0}}$\ for each mode $i$ can be expressed as $N=\left(  \frac
{k_{\parallel}^{2}+k_{\perp}^{2}.}{k_{\parallel}^{2}+f_{i}(k_{\perp}^{2}%
,B)}\right)  ^{1/2}.$ Depending on the momentum range, $N$ can be either
larger or smaller than unity. Anyway, below the first pair creation threshold
$0<k_{0}^{2}-k_{\parallel}^{2}<4m^{2}$ and, for larger energies, in the
regions adjacent from below to higher thresholds of creation of pairs
occupying excited Landau levels, $k_{0}^{2}-k_{\parallel}^{2}<\left(
m_{n}+m_{n^{\prime}}\right)  ^{2}$, the dispersion curves are
known\ \cite{Shabad1975}, \cite{Shabad2004} to lie outside the light cone,
$k^{2}=$ $k_{0}^{2}-\mathbf{k}^{2}>0,$ i.e. $f_{i}(k_{\perp}^{2})<k_{\perp
}^{2},$ hence\footnote{The inequality $N\lessgtr1$ also changes to the
opposite when poles of the polarization operator corresponding to creation by
a photon of mutually bound state of the electron-positron pair are crossed by
the dispersion curve\cite{ShabUs1985}.} $N>1$. This circumstance provokes an
(unsound) conclusion that ChR should exist in the vacuum.

However, causality implies that the group velocity of electromagnetic
radiation $\mathbf{v}_{\text{gr}}\mathbf{=}\frac{dk_{0}}{d\mathbf{k}}$ should
be smaller than 1
\begin{equation}
|\mathbf{v}_{\text{gr}}\mathbf{|}^{2}=\left(  \frac{\partial k_{0}}{\partial
k_{\parallel}}\right)  ^{2}+|\frac{\partial k_{0}}{\partial\mathbf{k}_{\perp}%
}|^{2}=\frac{k_{\parallel}^{2}+k_{\perp}^{2}\left(  f_{i}^{\prime}(k_{\perp
}^{2})\right)  ^{2}}{k_{0}^{2}}\leqslant1,\text{ \ \ \ \ \ \ }
\label{causality}%
\end{equation}
the prime indicates differentiation over $k_{\perp}^{2}$. Eq .$\left(
\ref{causality}\right)  $\ written in the form\ $k_{\perp}^{2}\left(
f_{i}^{\prime}(k_{\perp}^{2})\right)  ^{2}\leqslant k_{0}^{2}-k_{\parallel
}^{2}$ excludes the negativity of the difference $k_{0}^{2}-k_{\parallel}^{2}$
prescribed by the kinematical inequality (\ref{Domain}).\ Consequently, ChR
could only be a super-luminal tachyon, if dynamics of a theory admits its
appearance. \footnote{Note, that what matters for causality (\ref{causality}%
)\ is not the position of a dispersion curve inside or outside the light cone,
but its slope $\mid f_{i}^{\prime}(k_{\perp}^{2})\mid\leqslant\left(
\frac{k_{0}^{2}-k_{\parallel}^{2}}{k_{\perp}^{2}}\right)  ^{\frac{1}{2}}.$ The
right-hand side in this inequality is smaller than unity in the domains, where
the refractive index N exceeds unity (outside of the light cone), and it is
greater than unity otherwise .}

The above general conclusion that the photon dispersion curves do not get into
the domain (\ref{Domain}) is confirmed by all available calculations within
quantum electrodynamics of strong magnetic field, using both the polarization
tensor obtained by field-differentiation from the Euler-Heisenberg effective
Lagrangian (valid, according to \cite{Adorno}, in the infrared region under
the anisotropic conditions: $k_{0}^{2}-k_{\parallel}^{2}$ should be much less
than the inverse Compton length squared $m^{2},$ and $k_{\perp}^{2}$ much less
than the Larmour magnetic length$\sqrt{eB}$), or calculated as the one-loop
Furry diagram with the fermion propagators in it taken as solutions of the
Dirac equation with an external magnetic field \cite{Batalin}, \cite{Tsai},
\cite{Baier}, \cite{Shabad1975} (this result is not subject to any
frequency-momentum restrictions).

\subsection{Contrasting to the case of medium with the magnetic field}

When there is a homogenous medium, whose anisotropy is solely owing to the
magnetic field imposed on it, the Cherenkov radiation by an electron without
change of its Landau quantum number does occur along the same kinematical
lines as described in Subsect. 2.1. This time, however, the forbiddance
pointed in Subsect. 2.2 does not act, because of the additional violation of
the Lorentz invariance due to the medium, since its rest frame is specialized.
As an example of the needed medium let me refer to the electron-positron gas
at finite temperature and finite chemical potential placed in a magnetic
field.To achieve the Lorentz-covariant formulation of this system one should
introduce an extra, time-like, vector of four velocity of the medium $u_{\mu
},$ $u^{2}=1$. Now, apart from the invariants $k^{\mu}\widetilde{F}_{\mu\nu
}^{2}k^{\nu}$ and $k^{\mu}F_{\mu\nu}^{2}k^{\nu},$ the three eigenvalues of the
polarization tensor $\kappa_{i}$ may depend on other momentum-containing
Lorentz invariants: $\left(  uk\right)  ,u^{\mu}F_{\mu\nu}^{2}k^{\nu},u^{\mu
}\widetilde{F}_{\mu\nu}^{2}k^{\nu},u^{\mu}F_{\mu\nu}k^{\nu},u^{\mu
}\widetilde{F}_{\mu\nu}^{2}k^{\nu}.$ Then the form of the dispersion curves
$\left(  \ref{dspr}\right)  $ is no longer true and the principle of causality
cannot be given the form $\left(  \ref{causality}\right)  .$ Thereby the
forbiddance proof of ChR given above is invalidated.

When considering in \cite{Perez} the mechanisms of absorption of a photon via
creation of an electron-positron pair, both belonging to the plasma and
occupying the same Landau levels, $n=n^{\prime}$, it was found that the
antiHermitean parts of the polarization tensor, which are responsible for this
absorption, calculated using the temperature Green function method
\cite{Fradkin} within one-loop approximation, do exist in the domain $\left(
\ref{Domain}\right)  .$The calculated matrix element squared for this process
$\gamma\rightarrow e^{+}e^{-}$, indicated as inverse Cherenkov radiation in
\cite{Perez}, is the same as that for the process under consideration.

\subsection{Does the synchrotron radiation, $n\neq n^{\prime}$ include a
Cherenkov component?}

For $n\neq n^{\prime}$ equation $\left(  \ref{1}\right)  $%
\begin{equation}
p_{\parallel}=k_{\parallel}+p_{\parallel}^{\prime},\text{ \ \ \ \ }%
\varepsilon(p_{\parallel},n)=k_{0}+\varepsilon(p_{\parallel}^{\prime
},n^{\prime}),\text{ \ \ \ \ }p_{x}=\left(  \mathbf{k}_{\perp}\right)
_{x}+p_{x}^{\prime},
\end{equation}

\bigskip when squared, has two solutions%
\begin{equation}
p_{\parallel}^{(1,2)}=\frac{-k_{\parallel}J_{nn^{\prime}}\pm k_{0}%
\Lambda_{nn^{\prime}}}{2z}+k_{\parallel}=\frac{k_{\parallel}J_{n^{\prime}n}\pm
k_{0}\Lambda_{nn^{\prime}}}{2z}, \label{ResMom}%
\end{equation}
where $J_{nn^{\prime}}=z+2eB\left(  n^{\prime}-n\right)  $ \ and
\ $\Lambda_{nn^{\prime}}=\left(  J_{nn^{\prime}}^{2}+4m_{n}^{2}z\right)
^{\frac{1}{2}}$. In the limit $n=n^{\prime}$ the values $\left(
\ref{ResMom}\right)  $ coincide with $\left(  \ref{1}\right)  .$ In the domain%
\begin{equation}
-\left(  m_{n}-m_{n^{\prime}}\right)  ^{2}<z=k_{\parallel}^{2}-k_{0}^{2}<0
\label{negativezdomain}%
\end{equation}
\ both of them are solutions to equation $\left(  \ref{1}\right)  $
$n>n^{\prime}$, whereas equation $\left(  \ref{1}\right)  $ has no solutions
in this domain for $n<n^{\prime}.$ There is no solution in the adjacent domain
$-\left(  m_{n}-m_{n^{\prime}}\right)  ^{2}>z=k_{\parallel}^{2}-k_{0}^{2}$ in
either case. As for the domain $\left(  \ref{Domain}\right)  ,$ it is not
interesting as far as the vacuum is concerned, since as it was argued
previously, the photon dispersion curve does not get there, no matter whether
this is a Cherenkov photon nor a cyclotron photon. The energies of the
irradiating electron having the momenta $\left(  \ref{ResMom}\right)  $ are%
\[
\varepsilon^{\left(  1,2\right)  }=\frac{-k_{0}J_{nn^{\prime}}\pm
k_{\parallel}\Lambda_{nn^{\prime}}}{2z}+k_{0}=\frac{k_{0}J_{n^{\prime}n}\pm
k_{\parallel}\Lambda_{nn^{\prime}}}{2z}.
\]

\bigskip The corresponding speeds of the irradiating electrons are%
\begin{equation}
v_{\text{cync}}^{\left(  1,2\right)  }=\frac{p_{\parallel}^{\left(
1,2\right)  }}{\varepsilon^{\left(  1,2\right)  }}=\frac{k_{\parallel
}J_{n^{\prime}n}\pm k_{0}\Lambda_{nn^{\prime}}}{k_{0}J_{n^{\prime}n}\pm
k_{\parallel}\Lambda_{nn^{\prime}}}=\frac{-\lambda_{nn^{\prime}}N\cos\theta
\mp1}{-\lambda_{nn^{\prime}}\mp N\cos\theta}, \label{Cyclotron Speed}%
\end{equation}
where
\[
\lambda_{nn^{\prime}}=\frac{J_{n^{\prime}n}}{\Lambda_{nn^{\prime}}}%
=\frac{z+2eB\left(  n^{\prime}-n\right)  }{\left(  \left(  z+2eB\left(
n-n^{\prime}\right)  \right)  ^{2}+4m_{n}^{2}z\right)  ^{\frac{1}{2}}}<0.
\]

It holds that $\lambda_{nn^{\prime}}<0,$ since $n>n^{\prime}$ and $z<0$ in
(\ref{negativezdomain}). On the other hand $\left\vert z+2eB\left(  n^{\prime
}-n\right)  \right\vert >\left\vert z+2eB\left(  n-n^{\prime}\right)
\right\vert $. To decide, whether $\left\vert \lambda_{nn^{\prime}}\right\vert
$ is smaller or larger than unity, consider its values at the borders of the
domain $\left(  \ref{negativezdomain}\right)  $. Evidently at $z=0$ we have
$\lambda_{nn^{\prime}}=-1.$ At $z=-2\left(  m_{n}-m_{n^{\prime}}\right)  ^{2}$
the denominator is zero., hence $\lambda_{nn^{\prime}}=-\infty.$

To consider inequalities that the speed $\left(  \ref{CyclotronSpeed}\right)
$ of the synchrotron-radiating electron satisfies we start with the
inequality
\begin{equation}
-1>\lambda_{nn^{\prime}}>-\infty\label{Inequality for lambda}%
\end{equation}
and note that in the domain $\left(  \ref{negativezdomain}\right)  $ the ratio
$\frac{|k_{\parallel}|}{k_{0}}=N\left\vert \cos\theta\right\vert $\ lies
within the limits
\begin{equation}
\frac{\left\vert k_{\parallel}\right\vert }{\sqrt{\left(  m_{n}-m_{n^{\prime}%
}\right)  ^{2}+k_{\parallel}^{2}}}<N\left\vert \cos\theta\right\vert <1,
\label{inequality for Ncos}%
\end{equation}
therefore the refraction index\bigskip\ may be either larger or smaller than
unity for the cynchrotron radiation to be possible.

Bearing in mind that due to (\ref{inequality for Ncos}), \ $\left(
\ref{Inequality for lambda}\right)  $ the denominator in $\left(
\ref{Cyclotron Speed}\right)  $ is positive $-\lambda_{nn^{\prime}}\mp
N\cos\theta>0,$ the condition that the speed of the emitting electron should
not exceed that of light in the vacuum, $1\geqslant\left\vert v_{\text{sync}%
}\right\vert ,$ becomes$\ \lambda_{nn^{\prime}}\pm N\cos\theta\leqslant
-\lambda_{nn^{\prime}}N\cos\theta\mp1\leqslant-\lambda_{nn^{\prime}}\mp
N\cos\theta$. The right-hand inequality may be written as $\lambda
_{nn^{\prime}}(1-N\cos\theta)\leqslant\pm1\mp N\cos\theta$ , which is
equivalent to $\lambda_{nn^{\prime}}\leqslant\pm1$. This\ is fulfilled\ due to
\ $\left(  \ref{Inequality for lambda}\right)  .$\ The left-hand inequality
may be written as $\lambda_{nn^{\prime}}(1+N\cos\theta)\leqslant\mp1\mp
N\cos\theta$ , or $\lambda_{nn^{\prime}}\leqslant\mp1$ which is true again$.$
We conclude that the both radiating electrons are slower than light, as they
should be.

It is not demonstrated that the generic property of the Cherenkov radiation
that the emitting electron be faster than the phase speed of light $\left\vert
v_{\text{sync}}\right\vert \geqslant v_{\text{ph}}=\frac{1}{N}$ cannot be
fulfilled in any case. But there is no sign that the equality $\left\vert
v_{\text{sync}}\right\vert =v_{\text{ph}}$, even if possible, might be a
threshold for any new radiation.

We conclude that there is no Cherenkov part in the cyclotron radiation.

\section{\bigskip Conclusion}

Based on the classical theory of electromagnetic radiation, some authors have
recently come out with a statement, taken for granted also in a number of
subsequent works, that in an external magnetic field an electric charge moving
faster than the phase velocity of light $v_{\text{ph}}$ can generate
(Cherenkov) radiation because the vacuum with a background magnetic field
forms an optically dense medium with a refractive index greater than unity, so
that the phase velocity $v_{\text{ph}}$ = $N^{-1}$ is less than the speed of
light in vacuum, $v_{\text{ph}}$ $<1$. Although the last statement is a
well-known result of quantum electrodynamics, the process of radiation itself
in a quantum medium fixed in this way is considered in these works
classically, in particular, the trajectory of a charge in a magnetic field is classical.

On the contrary, the quantum kinetic analysis undertaken in our work showed
that Cherenkov radiation in a magnetic field in a vacuum does not take place.
An electron occupying a Landau quantum state and having a certain momentum
along the magnetic field creates synchrotron radiation, as it passes into a
state with a lower quantum Landau number, but its initial velocity may be both
greater and lesser than the phase velocity of light, so that the phase speed
is not a threshold for any radiation. In other words, in this radiation there
is no generic sign of Cherenkov radiation.

This statement applies to the radiation of an electron both in vacuum and in
an electron-positron plasma in a magnetic field --- it is a synchrotron radiation.

Cherenkov radiation could occur, according to kinematics, in a process with
conservation of the Landau quantum number in the electron-positron plasma.
However, in a vacuum, it is prohibited, as shown in the work, by the principle
of causality. This result is independent of the approximation used for the
photon polarization operator. The principle of causality requires that the
group velocity of a photon be limited to a value less than unity. This
requirement excludes the Cherenkov photon from falling into the region allowed
for it by the relativistic covariance in the external field. In the presence
of plasma, an additional violation of the Lorentz invariance arises in
comparison with that introduced by the external field, therefore the proof
presented in the work is destroyed and the prohibition on Cherenkov radiation
does not apply.

\section*{Acknowledgements}

Supported by RFBR project No. 20-02-00193.

\end{document}